\documentclass[preprint,1p,numbers,sort&compress,merge]{elsarticle}

\usepackage{lineno}
\usepackage{graphicx}
\usepackage{float}
\usepackage{bm}        % for math
\usepackage{amssymb}   % for math
\usepackage{slashed}   % for Dirac Slash
\usepackage{amsmath}   % for mutiline eqn
\usepackage{simplewick} % for contraction
\usepackage{verbatim}  % for multi-line comment
\usepackage{color} % for textcolor
\usepackage{xcolor} % for textcolor
\usepackage{appendix} % for appendix
\usepackage{subfig} % for sub fig
\usepackage{epstopdf}
\usepackage{hyperref}

\newcounter{YJC}

\graphicspath{{Figures/}}
\begin{document}
%\widetext

\title{Shining light on magnetic monopoles through high-energy muon colliders}
%\title{Search for magnetic monopoles at future high-energy muon colliders}

\author[1,2]{Ji-Chong Yang}

\author[1,2]{Yu-Chen Guo\corref{cor1}}
\ead{ycguo@lnnu.edu.cn}

\author[1,2]{Bing Liu}

\author[3]{Tong Li}

\cortext[cor1]{Corresponding author}

\address[1]{Department of Physics, Liaoning Normal University, Dalian 116029, China}
\address[2]{Center for Theoretical and Experimental High Energy Physics, Liaoning Normal University, Dalian 116029, China}
\address[3]{School of Physics, Nankai University, Tianjin 300071, China}

\begin{abstract}
The search for magnetic monopoles has been a longstanding concern of the physics community for nearly a century.
However, up to now, the existence of elementary magnetic monopoles remains an open question.
The electroweak 't Hooft-Polyakov monopoles have been predicted with mass at the order of 10 TeV.
This mass scale is unreachable at current colliders.
Recently, the muon colliders have gained much attention in the community due to technological developments.
The advantages of the muon beam encourage us to raise the high-energy option and consider the high-energy muon collider as a unique opportunity to search for magnetic monopoles.
This letter discusses the production of magnetic monopoles via the annihilation process and proposes the search for magnetic monopoles at future high-energy muon colliders.
\end{abstract}


\maketitle

%%%%%%%%%%%%%%%%%%%%%%%%%%%%%%%%%%%%%
\section{\label{sec1}Introduction}
%%%%%%%%%%%%%%%%%%%%%%%%%%%%%%%%%%%%%

% Presentation of the monopole
Magnetic monopoles (MMs) with magnetic charge were motivated by electric-magnetic symmetry.
Dirac first proposed that the existence of MMs could increase the symmetry of Maxwell's equations and explain the quantization of electric charge~\cite{Dirac:1931kp}.
%Dirac monopoles can exist in pure $U(1)$ gauge theory but are not required.
It was realized that any UV completion theory of an interacting $U(1)$ gauge field should contain MMs~\cite{CompletenessHypothesis}.
Recent research shows that the virtual effect of MMs can provide axion potential if the axion couples to the electric-magnetic gauge field~\cite{Fanjiji}.
MMs can also play as the role of dark matter candidate~\cite{darkcandidate}.
The monopole can be treated as a hypothetical point particle.
These point-like particles with unknown spin and mass carry a magnetic charge that is an integral multiple of the fundamental Dirac charge
\begin{equation}\label{diracrule}
g = \frac{e}{2\alpha_{\rm em}} N \,   = 68.5e \, N \equiv N \,   g_D\;,
\end{equation}
where $\alpha_{\rm em}\thickapprox1/137$ is the fine-structure constant, $N$ is an integer, $e$ is the electric charge, and $g_D = 68.5e$ is the fundamental Dirac charge.

% Mass range of 't Hooft-Polyakov monopoles
By extending the classical monopole with non-Abelian gauge theory by 't Hooft, Polyakov and others, composite monopoles naturally appear in the context of grand unified theories (GUT)~\cite{tHooft:1974kcl,Polyakov:1974ek}.
The mass of 't Hooft-Polyakov monopoles is determined by the scale of gauge symmetry breaking, which is of order $10^{16}$ GeV or higher for GUT-scale monopole. As a result, they cannot be produced in a realistic experiment.
Nevertheless, as some unification theories involve a number of symmetry-breaking scales, the electroweak symmetry-breaking can give rise to 't Hooft-Polyakov monopoles of mass at the order of $M_W/\alpha _{em}\sim 10~{\rm TeV}$ where $M_W$ is the $W$ boson mass~\cite{Lazarides:1980va,Kirkman:1981ck,Drukier:1981fq,Preskill:1984gd,Cho:1996qd,Kephart:2001ix}.
Except for the electroweak monopole of $5.5\;{\rm TeV}$ proposed by Ref.~\cite{Ellis:2016glu}, the $10\;{\rm TeV}$ monopole is not yet accessible to existing collider experiments. Thus, the heavy monopoles are mainly constrained by cosmological observations~\cite{MACRO:2002jdv,Balestra:2008ps,BAIKAL:2007kno,Hogan:2008sx,Abbasi:2010zz,ANITA-II:2010jck,Super-Kamiokande:2012tld,PierreAuger:2016imq,NOvA:2020qpg,IceCube:2021eye,ANTARES:2022zbr}, the analysis of moon rocks~\cite{Ross:1973it} or terrestrial materials~\cite{Kalbfleisch:2000iz,Kalbfleisch:2003yt} exposed to cosmologically relic monopoles~\cite{relicmm1}.

% Recent HEP Experiments: 1. ATLAS and MoEDAL  2. Schwinger mechanism (heavy-ion collisions) 3. Atmospheric Fixed Target Experiment
In recent years, composite MMs with masses as low as a few TeV have also been proposed in various cases~\cite{Baines:2018ltl,Mavromatos:2020gwk,Song:2021vpo}.
These theories offer the possibility of collider production of MMs, thus rekindling the enthusiasm of theorists and experimentalists in the search for MMs.
Recent searches for TeV-scale monopoles have been carried out by ATLAS~\cite{ATLAS:2019wkg} and MoEDAL~\cite{MoEDAL:2019ort} at the LHC.
These searches for MMs assume direct productions via the Drell-Yan/annihilation and Photon/Vector Boson Fusion (VBF) mechanism.
Another method of producing MMs is through the dual Schwinger effect~\cite{Schwinger:1951nm} in strong magnetic fields~\cite{Gould:2021bre}, which was performed by the MoEDAL experiment via Pb-Pb heavy-ion collisions at the LHC~\cite{MoEDAL:2021vix}.
In addition, recent studies proposed the production of MMs via collisions of cosmic rays bombarding the atmosphere and set leading robust bounds on the production cross-section of MMs in the $5-100$ TeV mass range~\cite{Iguro:2021xsu}.
It has been proposed that, the production and decay of monopolium bound state also provide sensitive probes of MMs at the LHC~\cite{monopolium1,monopolium2}.

% Importance of studying Monopoles at the Muon collider
In this letter, we investigate the production mechanisms of MMs at the muon collider.
This novel collider, operating at collision energies of 10-30 TeV~\cite{muoncollider1,muoncollider2,muoncollider4,Liu:2021jyc,Liu:2021akf}, would be able to produce 10 TeV monopoles.
Due to the large coupling constant between the photon and monopole, the interaction coupling is in the non-perturbative regime.
There is no established theory that allows for direct calculation of MMs production cross-section so far.
The electric-magnetic duality theory of Quantum Electrodynamics (QED) could be used as a basis to estimate monopole production cross-sections.
The common approach is to consider possible benchmark scenarios via tree-level Feynman-like diagrams.
The theoretical framework of this letter is based on electric-magnetic duality by considering effective field theories (EFTs) with only electromagnetic interactions.
Perturbative unitarity %~\cite{unitarityHistory1,unitarityHistory2,unitarityHistory3}
is taken into account.
The cross-section for production of monopole as well as the angular distribution are investigated.

%%%%%%%%%%%%%%%%%%%%%%%%%%%%%%%%%%%%%
\section{\label{sec2}Formalism for magnetic monopoles pair production}
%%%%%%%%%%%%%%%%%%%%%%%%%%%%%%%%%%%%%

\begin{figure}[htb]
\begin{center}
\includegraphics[width=0.3\linewidth]{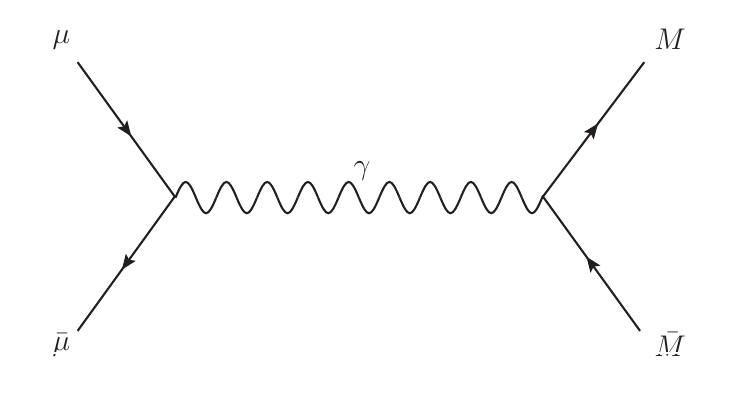}\\
\includegraphics[width=0.8\linewidth]{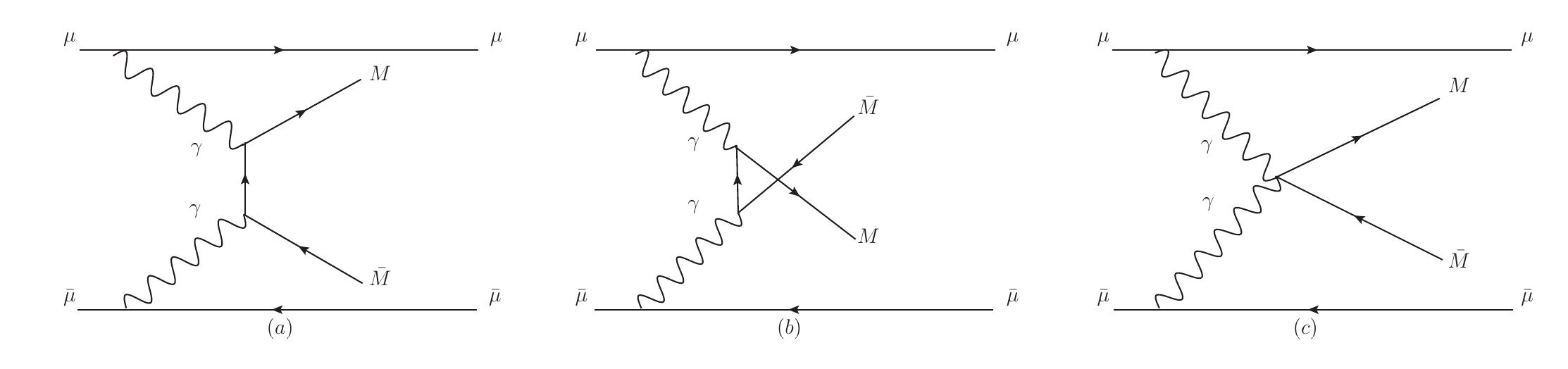}
\caption{Feynman-like diagrams for direct monopole pair production at leading order via the annihilation process (up); and VBF process (bottom $a,b,c$) at the muon collider. $M$ ($\bar{M}$) denotes the (anti-)MM.}
\label{fig:diagrams}
\end{center}
\end{figure}

The corresponding hypothesis is an effective $U(1)$ gauge theory obtained after appropriate dualisation of the pertinent field theories describing the interactions of MMs with photons.
By means of electric-magnetic duality, the charge and coupling of the MMs are often assumed to be velocity-dependent~(in the following, the velocity is denoted as $\beta$).
After replacing $e/v$ by $g/c$, the magnetic charge $g \, \frac{v}{c} \equiv g\, \beta $ would lead to the equivalence of the electron-monopole  scattering cross section with the corresponding Rutherford formula.
This replacement was widely used when discussing MMs pair production through annihilation and VBF processes~\cite{Kurochkin:2006jr,Dougall:2007tt,Epele:2012jn,Reis:2017rvb,Baines:2018ltl,Song:2021vpo,MoEDAL:2019ort,MoEDAL:2020pyb}.
In the following, we use $g(\beta)$ instead of $g$.
When the $\beta$-dependent interaction is considered, $g(\beta)=Ng_D\beta$, or otherwise, $g(\beta)=Ng_D$, where $N$ is the magnetic charge. The Lagrangians of the effective theories for the scalar, spinor and vector MMs are~\cite{Baines:2018ltl,monopolemodel}
\begin{equation}
\begin{split}
&\mathcal{L}^{S=0}=-\frac{1}{4}F^{\mu\nu}F_{\mu\nu}+(D^{\mu}\phi)^{\dagger}(D_{\mu}\phi)-M_s^2\phi^{\dagger}\phi\;,\\
&\mathcal{L}^{S=\frac{1}{2}}=-\frac{1}{4}F_{\mu\nu}F^{\mu\nu}+\overline{\psi}(i\slashed{D}-M_f)\psi-\frac{i}{4} g(\beta)\kappa_f F_{\mu\nu}\overline{\psi}[\gamma^{\mu},\gamma^{\nu}]\psi\;,\\
&\mathcal{L}^{S=1}=-\frac{1}{2}(\partial_{\mu}\mathcal{A}_{\nu})(\partial^{\nu}\mathcal{A}_{\mu})-\frac{1}{2}G_{\mu\nu}^{\dagger}G^{\mu\nu}-M_v^{2}W_{\mu}^{\dagger}W^{\mu}-ig(\beta)\kappa_v F^{\mu\nu}W^{\dagger}_{\mu}W_{\nu}\;,
\end{split}
\label{eq.2.1}
\end{equation}
where $\mathcal{A}_{\mu}$ is the photon field, $\phi$, $\psi$ and $W$ are the scalar, spinor and vector monopole fields with masses $M_{s,f,v}$, respectively, $F_{\mu\nu}=\partial_{\mu}\mathcal{A}_{\nu}-\partial_{\nu}\mathcal{A}_{\mu}$ is the electromagnetic field strength tensor, $D_{\mu}=\partial_{\mu}-ig (\beta)\mathcal{A}_{\mu}$ is the $U(1)$ covariant derivative, $G^{\mu\nu}=D^{\mu}W^{\nu}-D^{\nu}W^{\mu}$ provides the coupling of the magnetically charged vector field $W_{\mu}$ to the gauge field $\mathcal{A}_{\mu}$, and $\kappa$ is the interaction strength between the MMs and magnetic dipole moment.

At the muon collider, MMs pairs are produced by both the $\mu^+\mu^-$ annihilation and the VBF mechanism, as depicted in Fig.~\ref{fig:diagrams}.
For the $\mu^+\mu^-$ annihilation into MMs pairs, the cross-section would sharply peak when the invariant mass of MMs pair is close to the collision energy.
On the other hand, when the collision energy is much larger than the MMs pair mass, the VBF process dominates right after the threshold of MMs pair mass~\cite{Baines:2018ltl}.
Besides, in the latter case, the produced MMs typically have large velocities, possibly putting the unitarity into trouble.
Therefore, the golden channel to detect MMs depends on the center-of-mass energy as well as the mass range of MMs.

\begin{figure}[htb]
\begin{center}
\includegraphics[width=0.7\linewidth]{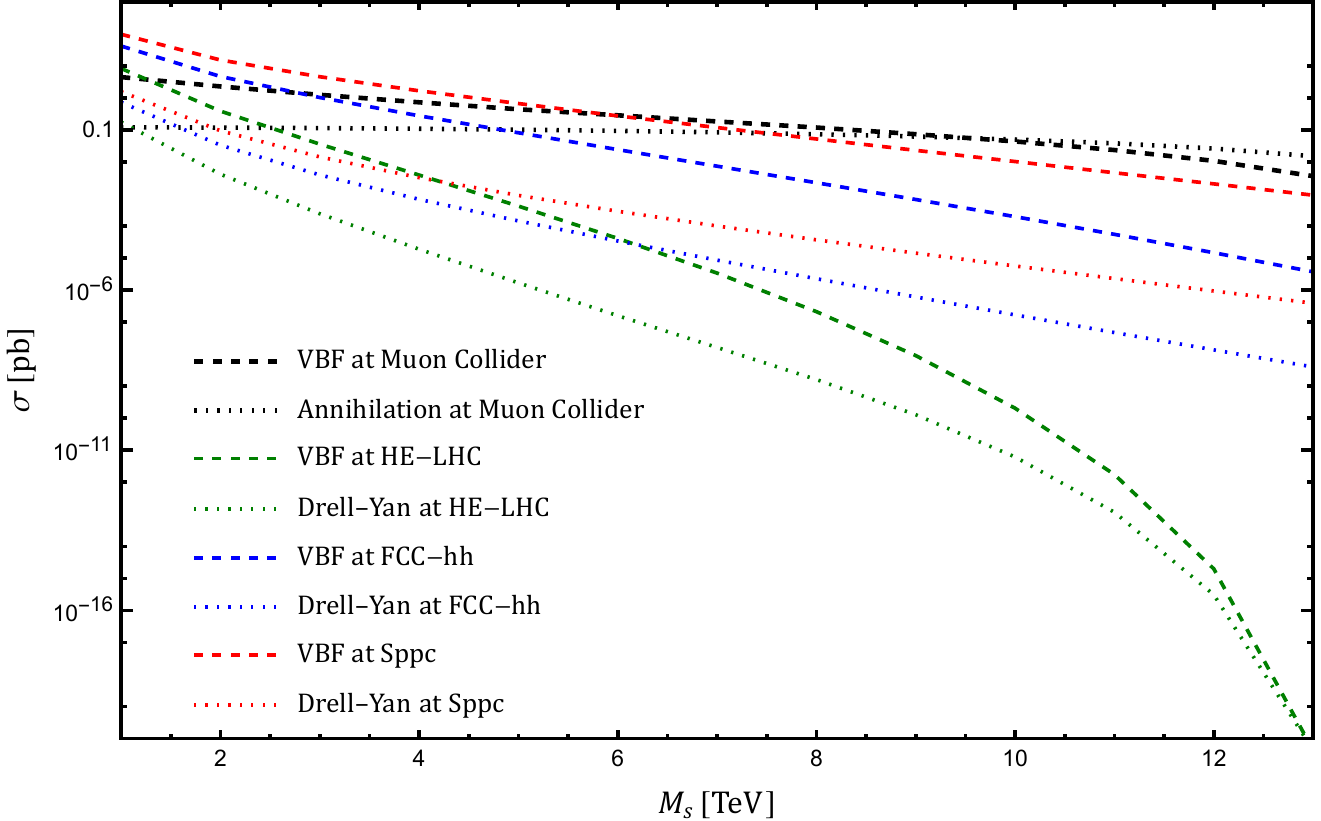}\\
\caption{The cross-sections of annihilation (or Drell-Yan) and VBF processes for scalar MMs, as a function of $M_s$ at 30 TeV muon collider, 27 TeV HE-LHC, 50 TeV FCC-hh and 75 TeV Sppc, respectively.
Here `VBF' represents $\mu^+\mu^-\to \ell^+\ell^- \phi\bar{\phi}$ or $pp\to j j \phi\bar{\phi}$ including the contributions with same final states but not containing a VBF subprocess.}
\label{fig:vbfannihilation}
\end{center}
\end{figure}

Taking scalar MM for illustration, the $\mu\mu$ annihilation cross-section becomes $\sigma\sim (1-4M_s^2/s)^{3/2}/s$ which is insensitive to the MM mass unless $M_s\approx \sqrt{s}/2$. By contrast, the VBF process is very sensitive to the MM mass due to the threshold suppression $\sim 1/M_s^2$.
When the mass of MM approaches $\sqrt{s}/2$, the annihilation process will again become dominant.
By using \verb"Madgraph_@aMC" toolkits~\cite{madgraph,feynrules,ufo}, the results are shown in Fig.~\ref{fig:vbfannihilation} for the scalar MM and $\beta$-independent coupling with $N=1$ at a $\sqrt{s}=30\;{\rm TeV}$ muon collider.
The kinematic features are studied using \verb"MLAnalysis"~\cite{Guo:2023nfu}. 

Since the purpose of this work is to illustrate the importance of the muon collider for a 't Hooft-Polyakov MM with mass of the order of $10$ TeV, we focus only on MMs with masses above $10$ TeV.
In this region, the $\mu\mu$ annihilation overcomes the VBF processes as shown above.
In the following, we will only consider the annihilation process.

At future $pp$ colliders, a majority of the partons reside in the range of energies much lower than the collision energy, leading to a small luminosity of the partons with large enough energy to produce the MMs.
The cross-sections at 27 TeV HE-LHC~\cite{HE-LHC}, 50 TeV FCC-hh~\cite{FCC-hh} and 75 TeV SppC~\cite{SppC} with the parton distribution function as \verb"NNPDF2.3"~\cite{NNPDF} are also shown in Fig.~\ref{fig:vbfannihilation}.
When $M_s=10\;{\rm TeV}$, the most optimistic case is the VBF at the Sppc, which is about $1/5$ of the annihilation at the muon collider.
One can conclude that the future high-energy muon collider has a unique advantage in detecting MMs with large masses.

%%%%%%%%%%%%%%%%%%%%%%%%%%%%%%%%%%%%%
\section{\label{sec3}Partial wave unitarity}
%%%%%%%%%%%%%%%%%%%%%%%%%%%%%%%%%%%%%

The coupling constant between the photon and monopole is too large to be valid for perturbation.
Note that, with a none-zero $\kappa_f$, the last term of $\mathcal{L}^{S=\frac{1}{2}}$ is a dimension-$5$ operator.
Meanwhile, it has been shown that, in the case of a vector MM, the unitarity is violated at large $s$ with $\kappa_v\neq 0$ for VBF processes.
Therefore, Eq.~(\ref{eq.2.1}) can only be viewed as an EFT.
Whether the EFT is valid in the energy range of interest needs to be checked before meaningful results can be obtained from tree level calculations.

The breakdown of perturbative theory can be determined by the violation of partial wave unitarity~\cite{partialwaveunitaritybound},%~\cite{partialwaveunitaritybound,jrr1},
which has been widely used in the study of SMEFT.%~\cite{Guo:2020lim,Fu:2021mub,Yang:2020slb,Guo:2021jdn,Yang:2021pcf,Yang:2022fhw,unitarity1,unitarity2,unitarity3,unitarity4,ubnew1,ubnew2,ubnew3,wprime}.
The $2\to 2$ helicity amplitudes $\mathcal{M}_{i\to f}(\theta,\phi)$ can be expanded as%~\cite{partialwaveexpansion}
\begin{equation}
\begin{split}
&\mathcal{M}_{i\to f}(\theta,\varphi)=8\pi\sum _J (2J+1)T_{i\to f}^{J,\lambda_1,\lambda_f}e^{i(\lambda_i-\lambda_f)\varphi}d^J_{\lambda_i,\lambda_f}(\theta,\varphi),
\end{split}
\label{eq.3.1}
\end{equation}
where $\varphi$ and $\theta$ are the azimuth and zenith angles of the MMs in the final state, $\lambda_{i,f}$ are helicity differences between the two particles in the initial and final states, $d^J_{\lambda_i,\lambda_f}(\theta,\varphi)$ are Wigner-D functions, and $T_{i\to f}^{J,\lambda_i,\lambda_f}$ are expansion coefficients
\begin{equation}
\begin{split}
&T_{i\to f}^{J,\lambda_i,\lambda_f}=\frac{1}{32\pi^2}\int d\varphi\int  d\cos (\theta) \mathcal{M}_{i\to f}(\theta,\varphi)e^{-i(\lambda_i-\lambda_f)\varphi}\left(d^J_{\lambda_i,\lambda_f}(\theta,\varphi)\right)^*\;.
\end{split}
\label{eq.3.2}
\end{equation}
In the case of inelastic scattering~\cite{unitarityWithMassive1},%~\cite{unitarityWithMassive1,unitarityWithMassive2},
\begin{equation}
\begin{split}
&\sum _{i\neq f}\beta _i \beta _f T_{i\to f}^{J,\lambda_i,\lambda_f}\leq 1,\;\;\;
 \beta_{i,f}=\frac{\sqrt{\left(s-(m_1+m_2)^2\right)\left(s-(m_1-m_2)^2\right)}}{s}\;,
\end{split}
\label{eq.3.3}
\end{equation}
where $m_{1,2}$ are masses of the two particles in the initial state or final state. Neglecting the masses of the particles in the initial state, one gets $\beta _i=1$. For the final state, $\beta_f=\beta=\sqrt{1-4M_{s,f,v}^2/s}$ is the velocity of the MM. The unitarity bound used in this letter is then $\beta \sum _{\lambda _i,\lambda _f} \left|T_{i\to f}^{J,\lambda_i,\lambda_f}\right|^2\leq 1$.

Considering the case of annihilation process $\mu^+\mu^-\to M\bar{M}$,% where $\bar{M}$ denotes the anti MM,
we find the unitarity bound as
$\beta ^3 e^2 g^2(\beta)/(144 \pi ^2)\leq 1$  for the case of scalar MMs. It leads to $\beta < \left(9/N^2\right)^{1/3}$ in the case of velocity independent interaction, and $\beta < \left(9/N^2\right)^{1/5}$ for the velocity dependent interaction. Therefore, the unitarity bound is generally satisfied for $N\leq 3$.
In the case of spinor MMs, the unitarity bound becomes
\begin{equation}
\begin{split}
&\frac{\beta e^2 g^2(\beta)}{72 \pi ^2 \left(1-\beta^2\right)} \left|\beta^4-4 \beta^2 (\kappa_f M_f-1) (2 \kappa_f M_f-1)+3 (1-2 \kappa_f M_f)^2\right|\leq 1.\\
\end{split}
\label{eq.3.4}
\end{equation}
For simplicity, we define a dimensionless parameter $\hat{\kappa}_f\equiv \kappa_f M_f$. For fixed $s$ and $N$, Eq.~(\ref{eq.3.6}) leads to the constraint on $\hat{\kappa}_f$ for a given $M_f$
\begin{equation}
\begin{split}
&\left|\hat{\kappa}_f-\frac{6 M_f^2}{8 M_f^2+s}\right|\leq \frac{M_f \sqrt{\frac{2 e^2 g^2(\beta) \left(4 M_f^2-s\right)^3+72 \pi ^2 s^{3/2} \left(8 M_f^2+s\right) \sqrt{s-4 M_f^2}}{1-\frac{4 M_f^2}{s}}}}{e g(\beta) s \left(8 M_f^2+s\right)},\\
\end{split}
\label{eq.3.5}
\end{equation}
note that, for $\beta$-dependent coupling, $g(\beta)$ is also a function of $M_f$.
For vector MMs, we have the unitarity bound as follows
\begin{equation}
\begin{split}
&\frac{\beta ^3 e^2 g^2(\beta)}{144 \pi ^2 \left(1-\beta^2\right)^2}\left|3 \beta^4-2 \beta^2 (2 \kappa_v (\kappa_v+3)+5)+4 \kappa_v (2 \kappa_v+3)+7\right|\leq 1.\\
\end{split}
\label{eq.3.6}
\end{equation}
A translation can be done similarly as Eq.~(\ref{eq.3.5}),
\begin{equation}
\begin{split}
&\left|\kappa _v +\frac{6M _v^2}{s+4M _v^2}\right|<\frac{2 M _v^2 \sqrt{\frac{144 \pi ^2 \sqrt{s} \left(4 M _v^2+s\right)}{\left(s-4 M _v^2\right)^{3/2}}-\frac{e^2 g^2(\beta) \left(12M _v^4-2 M _v^2 s+s^2\right)}{M _v^2 s}}}{eg(\beta) \left(4 M _v^2+s\right)}\\
\end{split}
\label{eq.3.7}
\end{equation}

\begin{figure}[htb]
\begin{center}
\includegraphics[width=0.415\linewidth]{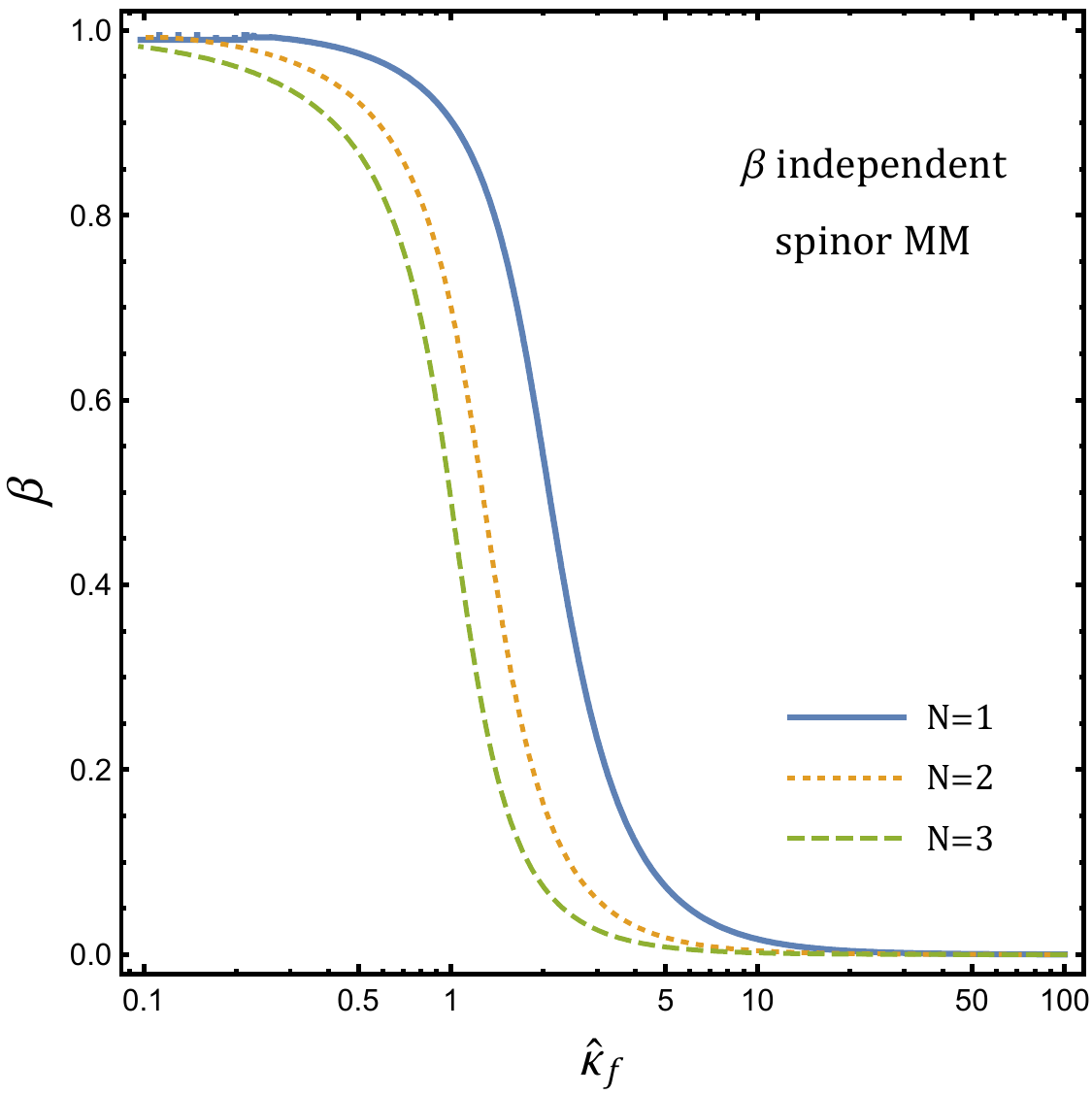}\quad
\includegraphics[width=0.39\linewidth]{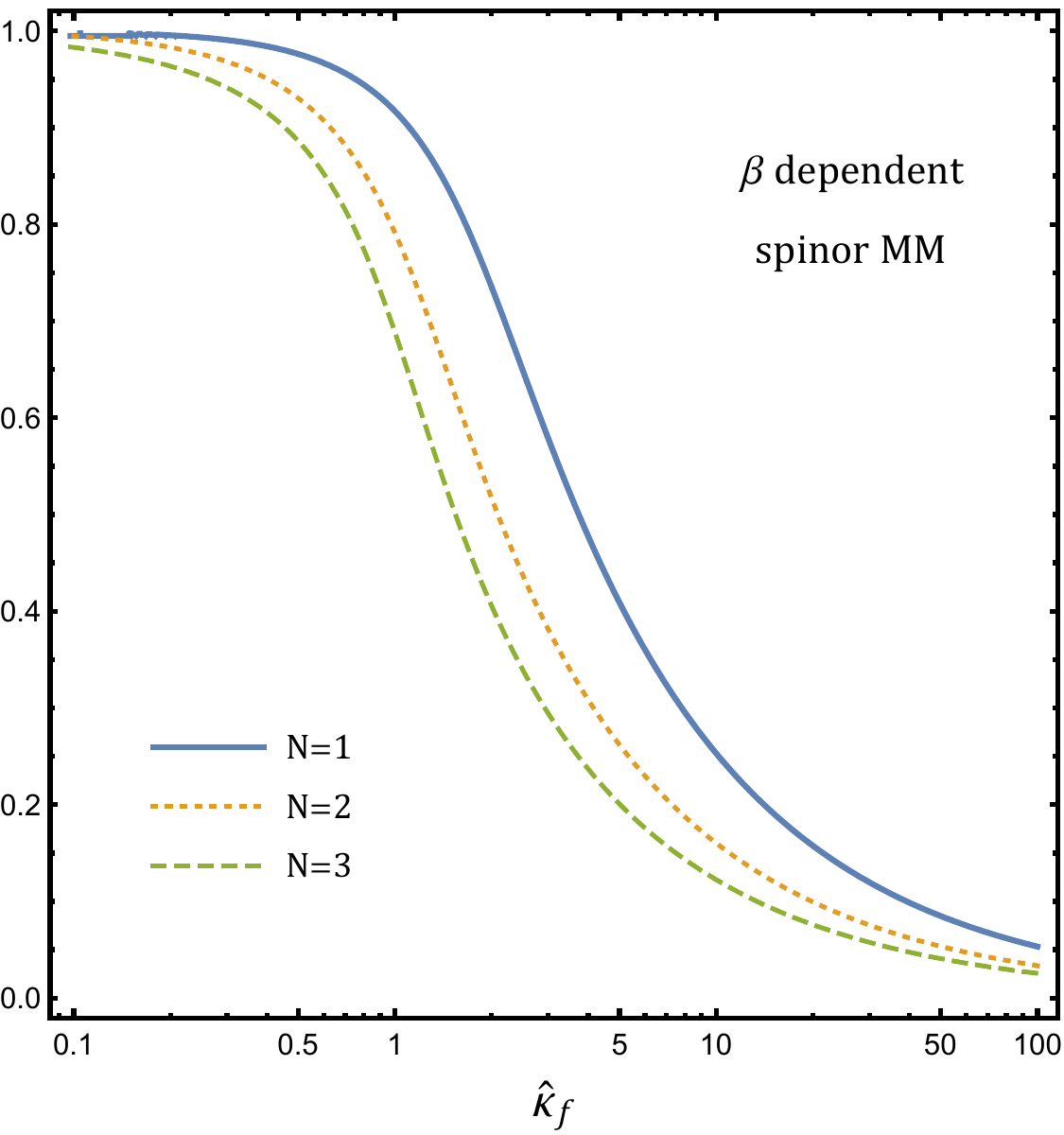}
\caption{The upper bound on $\beta$ for the case of a spinor MM. The cases of $\beta$-independent ($\beta$-dependent) $g$ are shown in the left (right) panel. The solid, dotted and dashed lines are for $N=1$, $2$ and $3$, respectively.}
\label{fig:betaf}
\end{center}
\end{figure}

\begin{figure}[htb]
\begin{center}
  \includegraphics[width=0.415\linewidth]{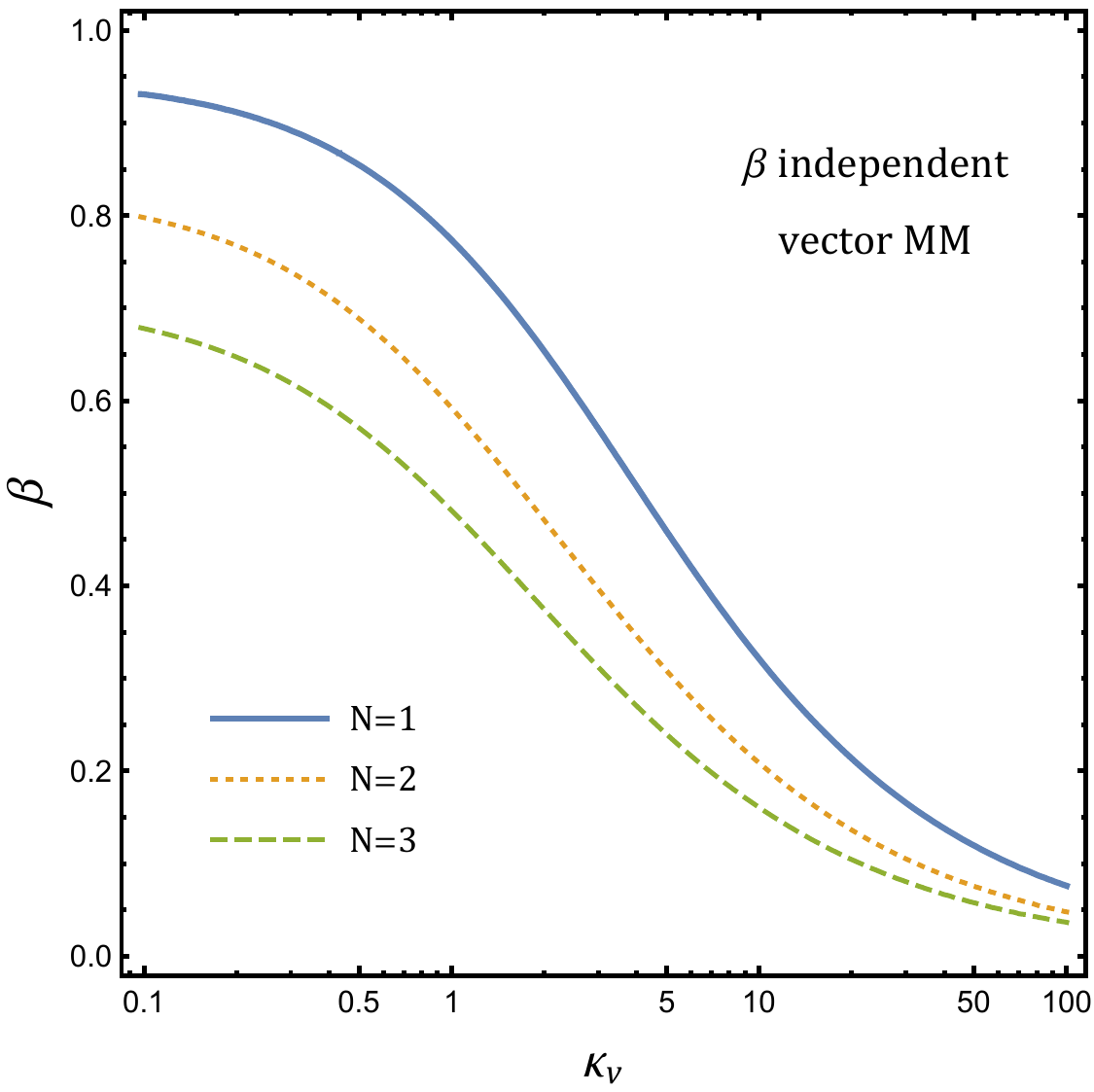}\quad
  \includegraphics[width=0.39\linewidth]{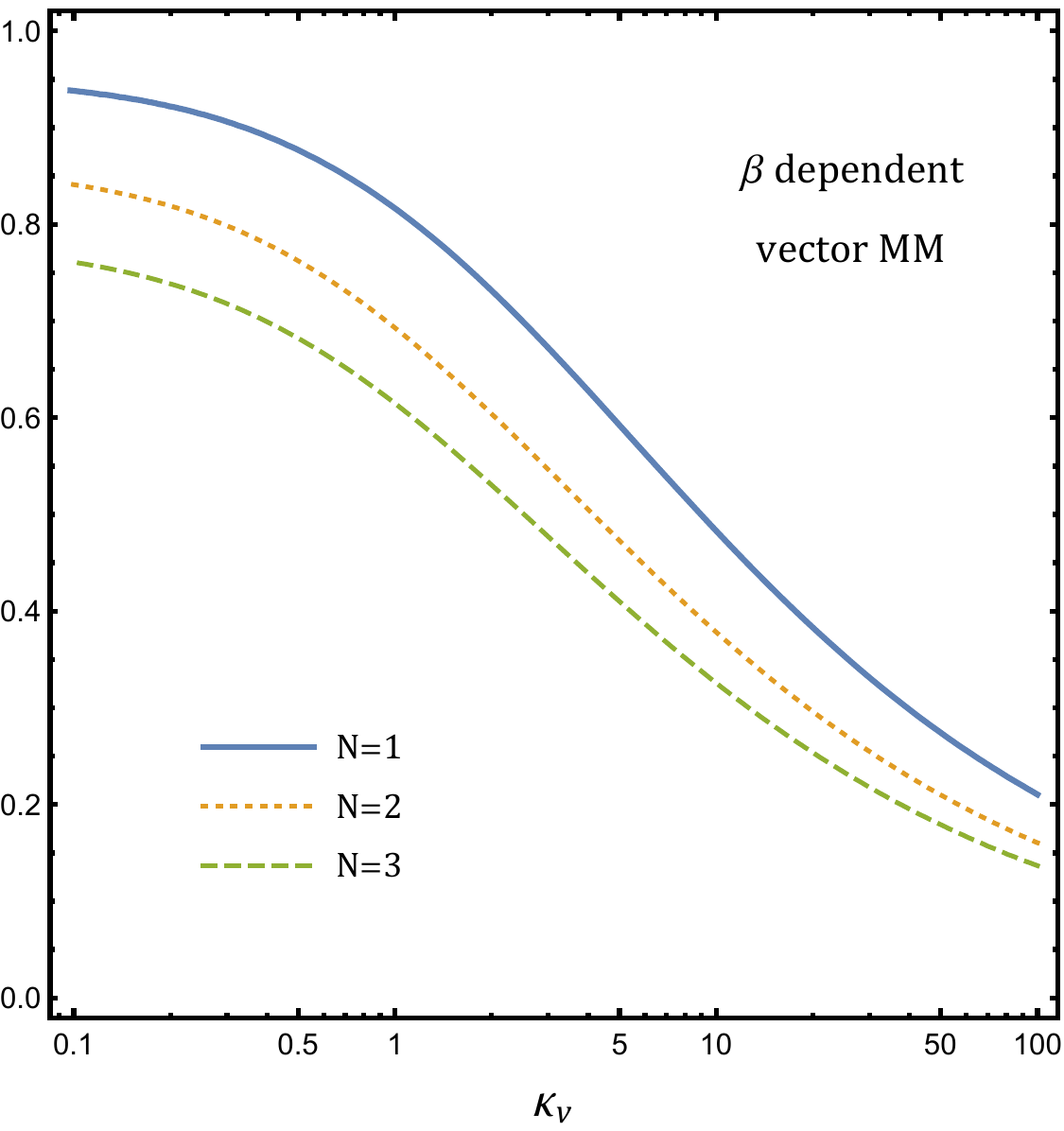}
  \caption{Same as Fig.~\ref{fig:betaf} but for the case of a vector MM.}
\label{fig:betav}
\end{center}
\end{figure}

Using Eqs.~(\ref{eq.3.4}) and (\ref{eq.3.6}), the upper bounds on $\beta$ are presented in Figs.~\ref{fig:betaf} and \ref{fig:betav}.
It can be seen that the allowed $\beta$ decreases rapidly along with increasing $\kappa_{f,v}$, for both spinor and vector MMs.
Since $\beta$ is a monotonic function of the MM mass, an upper bound on $\beta$ can be directly translated to a lower bound on $M_{f,v}$ for different $\sqrt{s}$.
Taking the $\beta$-independent coupling with $N=1$ as an example,
for the spinor MM case, the bound of $\beta<0.234$ when $\hat{\kappa}_f=3$ results in $0.5\sqrt{s}>M_f>0.486\sqrt{s}$~($15\;{\rm TeV}>M_f>14.6\;{\rm TeV}$ at $\sqrt{s}=30 \; {\rm TeV}$).
For the vector MM case, $\beta<0.313$ when $\kappa_v=10$ gives $0.5\sqrt{s}>M_v>0.444\sqrt{s}$~($15\;{\rm TeV}>M_f>14.2\;{\rm TeV}$ at $\sqrt{s}=30 \; {\rm TeV}$).
It turns out that a large $\kappa$ leaves a small mass window to detect the MM. When the mass of MM is lower than this bound, the EFT no longer succeeds to describe the reactions perturbatively. In the following studies, the results are presented with unitarity taken into account.

%%%%%%%%%%%%%%%%%%%%%%%%%%%%%%%%%%%%%%%%%
\section{\label{sec4} Monopole production in muon collisions}
%%%%%%%%%%%%%%%%%%%%%%%%%%%%%%%%%%%%%%%%

\begin{figure}[htb]
\begin{center}
\includegraphics[width=0.46\linewidth]{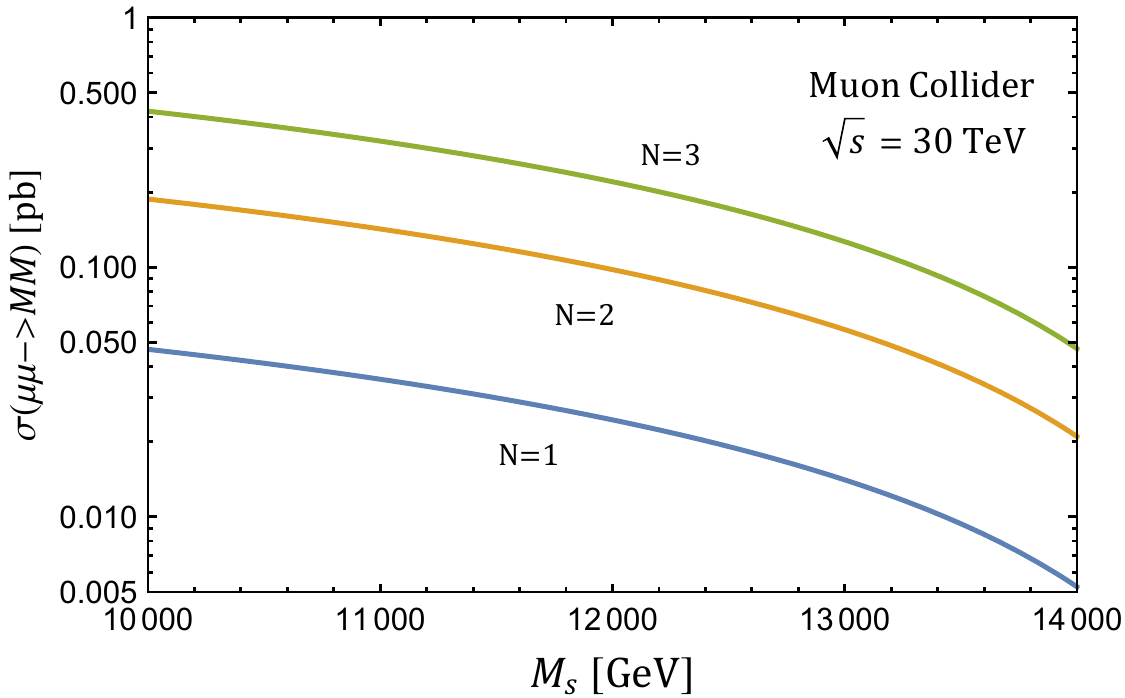}
\includegraphics[width=0.435\linewidth]{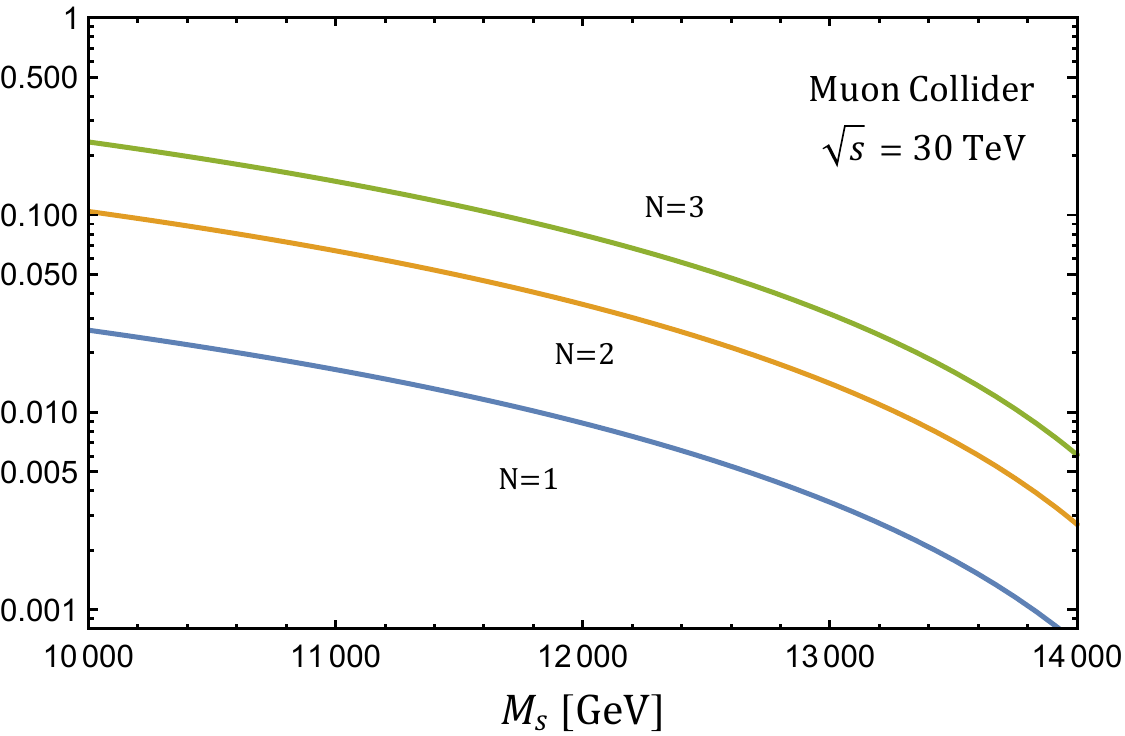}
\caption{$\sigma _s$ as a function of $M_s$ for $N=1,2,3$. The left (right) panel corresponds to the $\beta$-independent ($\beta$-dependent) coupling.}
\label{fig:crosss}
\end{center}
\end{figure}

\begin{figure}[htb]
\begin{center}
\includegraphics[width=0.48\linewidth]{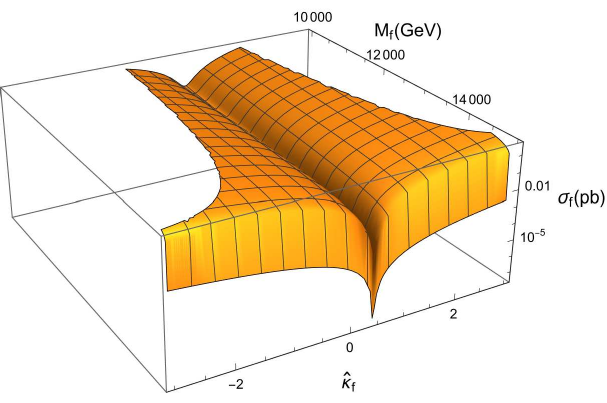}
\includegraphics[width=0.48\linewidth]{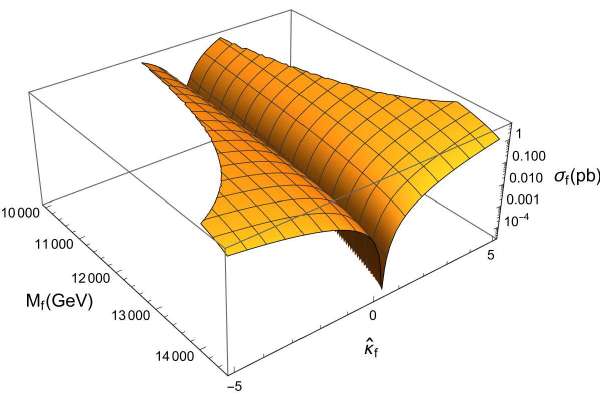}\\
\includegraphics[width=0.48\linewidth]{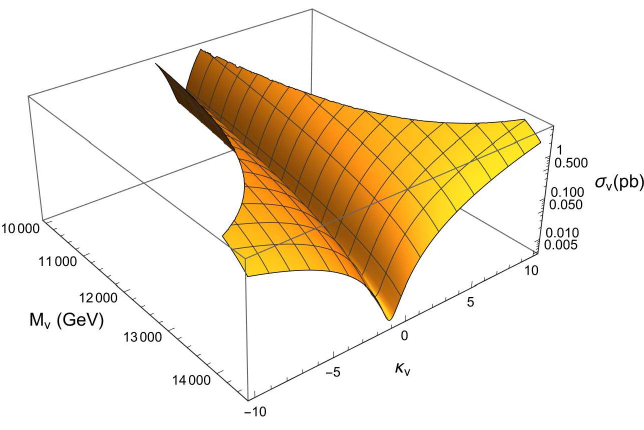}
\includegraphics[width=0.48\linewidth]{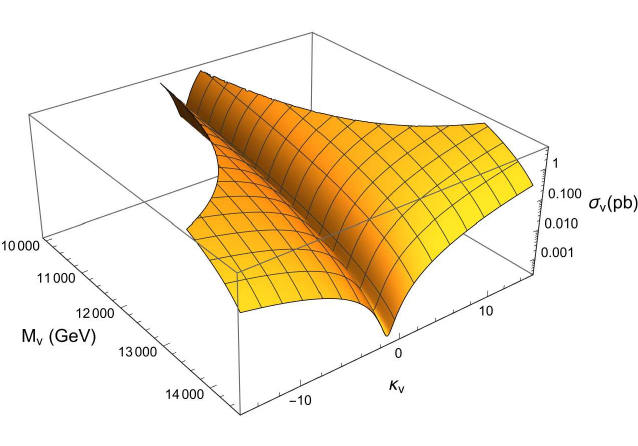}\\
\caption{$\sigma _{f,v}$ as functions of $M_{f,v}$ and $\hat{\kappa}_f$ or $\kappa_{v}$. For illustration, only $N=1$ case is shown. The top (bottom) panels correspond to $\sigma _f$ ($\sigma_v$) and $\hat{\kappa}_f$ ($\kappa_v$).
The left (right) panels are for the $\beta$-independent ($\beta$-dependent) coupling.
The regions that are forbidden by unitarity bounds are cropped out.
}
\label{fig:crossfv}
\end{center}
\end{figure}

In this section, we evaluate the production rate of MM pairs at future muon colliders.
%Since detectors usually do not cover the entire zenith angle, we consider the cross-section as
%\begin{equation}
%\begin{split}
%&\sigma = \int _{\Delta \theta}^{\pi-\Delta %\theta}\frac{d\sigma}{d \theta} d  \theta,
%\end{split}
%\label{eq.4.0}
%\end{equation}
%where $\Delta \theta$ is the minimal zenith angle that can be covered.
%
%\TL{removed the $\sigma$ formula}
Taking $\Delta \theta$ as the minimal zenith angle that can be covered, the cross-sections of MM productions via the annihilation process are
\begin{equation}
\begin{split}
&\sigma _s = \frac{e^2 g^2(\beta) \left(9 \cos (\Delta \theta)-\cos (3 \Delta \theta)\right) \left(s-4 M_s^2\right)^{3/2}}{384 \pi  s^{5/2}},\\
&\sigma _f =\frac{e^2 g^2(\beta) \cos (\Delta \theta) \sqrt{s-4 M_f^2}}{96 \pi  M_f^2 s^{5/2}} \left(-\cos (2 \Delta \theta) \left(4 M_f^2-s\right) \left(M_f^2-\hat{\kappa}_f^2 s\right)\right.\\
&\left.+5 \hat{\kappa}_f^2 s^2+(4 \hat{\kappa}_f (7 \hat{\kappa}_f-12)+7) M_f^2 s+20 M_f^4\right),\\
&\sigma _v =-\frac{e^2 g^2(\beta) \cos (\Delta \theta) \left(s-4 M_v^2\right)^{3/2} }{768 \pi  M_v^4 s^{5/2}}\left(\cos (2 \Delta \theta) \left(-2 \left(\kappa_v^2+1\right) M_v^2 s+\kappa_v^2 s^2+12 M_v^4\right)\right.\\
&\left.-5 \kappa_v^2 s^2-2 (\kappa_v (7 \kappa_v+24)+7) M_v^2 s-60 M_v^4\right).
\end{split}
\label{eq.4.1}
\end{equation}
One can see that, for a small $\Delta \theta$, $\sigma _{s,f,v}$ are insensitive to $\Delta \theta$. Therefore, in numerical results, we take $\Delta \theta = 0$.
We consider MMs with masses of the order of $10$ TeV produced at muon collider with $\sqrt{s}=30$ TeV.
At this c.m. energy, the cross-sections as functions of monopole mass and magnetic dipole moment are shown in Figs.~\ref{fig:crosss} for scalar MM and \ref{fig:crossfv} for spinor and vector MM.
When $\hat{\kappa}_f=6 M_f^2/\left(8M_f^2+s\right)$ and $\kappa_v=-6 M_v^2/\left(4 M_v^2+s\right)$, $\sigma _f$ and $\sigma _v$ reach the minima, respectively.
The minima are
\begin{equation}
\begin{split}
&\sigma _f^{\rm min} =\frac{e^2 g^2(\beta) \left(s-4 M_f^2\right)^{5/2}}{12 \pi  s^{5/2} \left(8 M_f^2+s\right)},\;\;
\sigma _v^{\rm min} = \frac{e^2 g^2(\beta)  \left(s-4 M_v^2\right)^{3/2} \left(12 M_v^4-2 M_v^2 s+s^2\right)}{48 \pi  M_v^2 s^{5/2} \left(4 M_v^2+s\right)}.
\end{split}
\label{eq.4.2}
\end{equation}

For $N=1$ and $\beta$-dependent couplings, in the mass range of $10\;{\rm TeV}<M_{s,f,v}<14\;{\rm TeV}$, we obtain $26.06\;{\rm fb}>\sigma _s > 0.68\;{\rm fb}$, $30.65\;{\rm fb}>\sigma _f^{\rm min} > 0.13\;{\rm fb}$ and $150.33\;{\rm fb}>\sigma _v^{\rm min} > 1.88\;{\rm fb}$.
The designed luminosity for muon colliders is expected to be $10\sim 90\;{\rm ab}^{-1}$~\cite{muoncollider4}.
Taking $\sigma _{s}$ or $\sigma_{f,v}^{\rm min}=10\;{\rm ab}$ as a fiducial cross-section, we find the reachable masses are $M_s=14.82$ TeV, $M_f=14.52$ TeV and $M_v=14.88$ TeV, respectively.

The angular distribution of the MM production is useful information for detecting the MMs.
The results are independent of $g(\beta)$ and are given by
\begin{equation}
\begin{split}
&\frac{1}{\sigma _s}\frac{d\sigma _s}{d \cos \theta} =\frac{3}{4} \left(1-\cos ^2 (\theta)\right),\;\;\;\;
 \frac{1}{\sigma _{f,v}}\frac{d\sigma _{f,v}}{d \cos \theta} =\frac{1}{N_{f,v}} \left(1+ A_{f,v}\cos (2\theta)\right),\\
&N_f=\frac{8 \left(2 \left(4 \hat{\kappa}_f^2-6 \hat{\kappa}_f+1\right) M_f^2 s+\hat{\kappa}_f^2 s^2+4 M_f^4\right)}{3 \left(\hat{\kappa}_f^2 s^2+(4 \hat{\kappa}_f (3 \hat{\kappa}_f-4)+3) M_f^2 s+4 M_f^4\right)}, A_f=\frac{\left(s-4 M_f^2\right) \left(M_f^2-\hat{\kappa}_f^2 s\right)}{\hat{\kappa}_f^2 s^2+(4 \hat{\kappa}_f (3 \hat{\kappa}_f-4)+3) M_f^2 s+4 M_f^4},\\
&N_v=\frac{8 \kappa_v^2 s^2+32 (\kappa_v (\kappa_v+3)+1) M_v^2 s+96 M_v^4}{3 \kappa_v^2 s^2+6 (\kappa_v (3 \kappa_v+8)+3) M_v^2 s+36 M_v^4}, A_v=\frac{2 \left(\kappa_v^2+1\right) M_v^2 s-\kappa_v^2 s^2-12 M_v^4}{\kappa_v^2 s^2+2 (\kappa_v (3 \kappa_v+8)+3) M_v^2 s+12 M_v^4}.\\
\end{split}
\label{eq.4.3}
\end{equation}
The scalar MMs are dominantly produced along the direction perpendicular to the beam line direction.
For the spinor MM and vector MM, the production direction depends on the sign of $A_{f,v}$.
%$A_{f,v}$ as functions of $M_{f,v}$ and $\hat{\kappa}_f$ or $\kappa_v$ are shown in Fig.~\ref{fig:afv}.
%
For the spinor MMs, $A_f<0$ when $|\hat{\kappa}_f|<M_f/\sqrt{s}$ and they are produced along the perpendicular direction. Otherwise, one has $A_f>0$ and the spinor MMs are produced along the beam line direction.
When $\hat{\kappa}_f =1/2$, $A_f$ reaches the minimum and $d\sigma _f/\left(\sigma _f d \cos \theta\right) =3\left(1-\cos ^2 (\theta)\right)/4$. When $\hat{\kappa}_f =2M_f^2/s$, $A_f$ reaches the maximum and $d\sigma _f/\left(\sigma _f d \cos \theta\right) =3\left(3+\cos  (2\theta)\right)/16$.
For the vector MMs, when $M_v<\sqrt{s/6}$~($12.25\;{\rm TeV}$ at $\sqrt{s}=30\;{\rm TeV}$) and $| \kappa _v |< \sqrt{\left(2M_v^2(s-6M_v^2)\right)/\left(s(s-2M_v^2)\right)}$, one has $A_v>0$ and the MMs are dominantly produced along the beam line direction. Otherwise, the MMs are produced along the perpendicular direction for  $A_v<0$.
When $\kappa _v=-1$, $A_v$ reaches the minimum and $d\sigma _v/\left(\sigma _v d \cos \theta\right) =3\left(1-\cos ^2 (\theta)\right)/4$.
When $\kappa _v=-2 \left(6 M_v^4-M_v^2 s\right)/\left(s \left(2 M_v^2-s\right)\right)$, $A_v$ reaches the maximum and one gets
\begin{equation}
\begin{split}
&\frac{1}{\sigma _{v}}\frac{d\sigma _{v}}{d \cos \theta} =\frac{3 \left(36 M_v^4-8 M_v^2 s+\left(12 M_v^4-8 M_v^2 s+s^2\right) \cos (2 \theta)+3 s^2\right)}{16 \left(12 M_v^4-2 M_v^2 s+s^2\right)}.
\end{split}
\label{eq.4.4}
\end{equation}

%Various techniques have been used at the colliders to detect MMs based on different theoretical assumptions.
%The CDF experiment designed a special trigger based on the relativistic effects that MMs are accelerated along the uniform solenoidal magnetic field in a parabola slightly distorted~\cite{CDF:2005cvf}.
%The position, time and the photomultiplier tube pulses of the trigger can be measured by the central outer tracker (COT) and the time-of-flight (TOF) detector.
%The MMs produced at \reviseyjc{the} LHC exhibit characteristics of the long-lived highly ionizing particles which would quickly slow down and get trapped in the material surrounding the interaction points.
%The trapped MMs would produce a large number of high-threshold hits and a large number of $\delta$ rays emitted from the material~\cite{Ahlen:1978jy,Ahlen:1980xr}.
%ATLAS measured the ionization energy loss of the magnetic monopole as the monopole signature by the transition radiation tracker (TRT) in the inner detector and the liquid argon (LAr) sampling electromagnetic  calorimeter (EMC).
%Based on the assumption that monopole binding to matter through interactions with atoms or nuclei is strong, MoEDAL~\cite{MoEDAL:2021vix} searched for MMs by looking for induced persistent currents after passage through a superconducting magnetometer~\cite{Burdin:2014xma}. Whatever the assumptions are, the measurements in LHC experiments are determined by the ionization energy loss of the MMs which could be velocity dependent, when the MMs pass through the material~\cite{Ahlen:1978jy,Ahlen:1980xr}.

Various techniques have been used at the colliders to detect MMs based on different theoretical assumptions.
The CDF experiment designed a special trigger based on the relativistic effects that MMs are accelerated along the uniform solenoidal magnetic field in a parabola slightly distorted~\cite{CDF:2005cvf}.
%The position, time and the photomultiplier tube pulses of the trigger can be measured by the central outer tracker (COT) and the time-of-flight (TOF) detector.
The MMs produced at the LHC exhibit characteristics of the long-lived highly ionizing particles which would quickly slow down and get trapped in the material surrounding the interaction points and produce a large number of high-threshold hits and a large number of $\delta$ rays emitted from the material~\cite{Ahlen:1978jy,Ahlen:1980xr}.
%ATLAS measured the ionization energy loss of the magnetic monopole as the monopole signature by the transition radiation tracker (TRT) in the inner detector and the liquid argon (LAr) sampling electromagnetic  calorimeter (EMC).
Based on the assumption that monopole are strongly bound to aluminum nuclei, MoEDAL~\cite{MoEDAL:2021vix} searched for MMs by looking for induced persistent currents after passage through a superconducting magnetometer~\cite{Burdin:2014xma}. Whatever the assumptions are, the measurements at the LHC are determined by the ionization energy loss of the MMs which could be velocity dependent~\cite{Ahlen:1978jy,Ahlen:1980xr}.

%%%%%%%%%%%%%%%%%%%%%%%%%%%%%%%%%%
\section{\label{sec5}Summary}
%%%%%%%%%%%%%%%%%%%%%%%%%%%%%%%%%%

In this letter, we investigate the capability of the future muon colliders to detect the MMs with the mass of the order of $10$ TeV in the framework of EFTs.
We find that the annihilation process overcomes the VBF processes when probing a MM with mass close to $\sqrt{s}/2$.
Since the coupling constant is not in a perturbative region, the tree-level partial wave unitarity bounds are discussed, which result in upper bounds on the integral multiple of the fundamental Dirac charge $N$, and lower bounds on the mass of MMs, or upper bounds on magnetic dipole moments $\kappa _{f,v}$.

The cross-sections of the MM pair production via the annihilation process are calculated at the muon collider.
%For $N=1$ and $\beta$-dependent couiplings, the cross-sections that the muon collider can reach are $0.68-26.06$ fb for scalar MM, $0.13-30.65$ fb for spinor MM, and $1.88-150.33$ fb for vector MM when $10\;{\rm TeV}<M_{s,f,v}<14\;{\rm TeV}$.
Taking a fiducial cross-section of $10$ ab, the reachable masses are $M_s=14.82$, $M_f=14.52$ and $M_v=14.88\;{\rm TeV}$, respectively.
Compared with the LHC, the muon colliders provide a unique opportunity of the search for electroweak 't Hooft-Polyakov MMs.
The angular distributions of the MM production are also investigated.
Generally, the scalar MM, the spinor MM with a small $|\hat{\kappa}_v|$ or the vector MM with a large $M_v$ can be produced along the perpendicular direction.

Finally, inspired by the measuring methods in experiments at LHC, we propose that the MMs detectors should be designed for the future muon collider because it provides us the first possible source of 't Hooft-Polyakov MMs at colliders.

\section*{Acknowledgment}

\noindent
This work was supported in part by the National Natural Science Foundation of China under Grants Nos. 11905093 and 12147214, the Natural Science Foundation of the Liaoning Scientific Committee No.~LJKZ0978 and the Outstanding Research Cultivation Program of Liaoning Normal University (No.21GDL004).
T.L. is supported by the National Natural Science Foundation of China (Grants No. 11975129, 12035008) and ``the Fundamental Research Funds for the Central Universities'', Nankai University (Grant No. 63196013).

\bibliography{MM}
\bibliographystyle{elsarticle-num}

\end{document}